\documentclass[usenatbib]{mn2e}
\input epsf
\pdfoutput=1
\usepackage{graphicx}

\newcommand{\beq}{\begin{equation}}
\newcommand{\eeq}{\end{equation}}
\newcommand{\beqary}{\begin{eqnarray}}
\newcommand{\eeqary}{\end{eqnarray}}

\title[PWN G16.73+0.08/SNR G16.7+0.1]{{\it Chandra} Observation of PWN G16.73+0.08 in SNR G16.7+0.1} 
\author[Chang, Chung, Yang and Tian]{H.-K. Chang$^{1,2}$\thanks{E-mail: hkchang@phys.nthu.edu.tw}, 
S.-F. Chung$^1$, C.-Y. Yang$^1$ and W. W. Tian$^{3,4}$
\\
$^{1}$Institute of Astronomy, National Tsing Hua University, 
Hsinchu 30013, Taiwan\\ 
$^{2}$Department of Physics, National Tsing Hua University, 
Hsinchu 30013, Taiwan\\
$^{3}$Key Laboratory of Optical Astronomy, National Astronomical Observatories, Chinese Academy of Sciences, 
Beijing 100049, China\\
$^{4}$Department of Physics and Astronomy, University of Calgary, 
Calgary, Alberta T2N 1N4, Canada
} 

\begin{document}

\date{Accepted 2017 November 13 . Received 2017 November 13 ; in original form 2017 June 25}

\pagerange{\pageref{firstpage}--\pageref{lastpage}} \pubyear{2017}

\maketitle

\label{firstpage}

\begin{abstract}
We present 
X-ray observations
of PWN G16.73+0.08/SNR G16.7+0.1 using archival data
of {\it Chandra} ACIS.
The X-ray emission peak location of 
this pulsar wind nebula is found to be offset by 24 arcsec from the centre of the 1.4-GHz emission of this nebula.
The X-ray nebula is elongated in the direction from the X-ray peak to the 1.4-GHz emission centre.
This offset suggests that G16.73+0.08 is an evolved pulsar wind nebula interacting with the supernova remnant reverse shock. 
We 
identify
a point source, CXO J182058.16-142001.5, near the location of the X-ray peak.
The spectrum of the X-ray nebula can be described by an absorbed power law of photon index $0.98^{+0.79}_{-0.71}$
and hydrogen column density $N_{\rm H}=4.99^{+2.75}_{-2.28}\times 10^{22}$ cm$^{-2}$.  
CXO J182058.16-142001.5 is likely a pulsar. We estimate its spin-down power to be about
$2.6\times 10 ^{36}$ erg s$^{-1}$. Assuming its age at 3000 and 10,000 years, its
dipole magnetic field strength at the polar surface is estimated to be about
$4.2 \times 10^{13}$ G and $1.3 \times 10^{13}$ G, respectively.  
\end{abstract}


\section{Introduction}

The term 'Pulsar Wind Nebulae' (PWNe) was first proposed by \citet{hester88a}, motivated by
theoretical and observational works in the 1980's on the interpretation of the 
observed radio and X-ray nebulae or cores in supernova remnants (SNRs). 
It was realized that these objects are not part of remnants of supernova explosion events, 
but objects driven by the relativistic winds injected from the central pulsar 
and interacting with ISM or with supernova ejecta
(e.g., \citet{kennel84,michel85,chevalier87,kulkarni88,hester88b}). 
The study of PWNe has provided information about physical properties of ISM and supernova ejecta,
about the composition, energetics and evolution of the winds, and about the wind injection rate of the pulsar.
In addition, pulsar winds are likely related to the so-called positron excess in cosmic rays, which otherwise may
have a dark-matter origin (e.g., \citet{erlykin13,feng16}).
PWNe are also found to be major TeV sources in the sky.
To date, about 60 PWNe are found, although of nearly half of which the central pulsars are yet to be identified.
For a more comprehensive review, readers are referred to, e.g., 
\citet{kargaltsev15}, \citet{porth17}, and \citet{reynolds17}.

G16.73+0.08 is believed to be a PWN in the SNR G16.7+0.1.
Its radio emission from 90 cm to 2 cm was reported in \citet{helfand89}.
{\it ASCA} observation yielded a detection of an unresolved source \citep{sugizaki01}.
Later in {\it XMM} observation 
an extended source was revealed, which is located near the geometric centre of SNR G16.7+0.1 \citep{helfand03}.
This gives a hint of possible location offset between the {\it XMM}-observed X-ray nebula and the radio nebula 
of G16.73+0.08, because the radio nebula is about 28 arcsec away from the geometric centre of G16.7+0.1.
We noticed that {\it Chandra} had also observed this source in 2003 
and therefore started the analysis of this {\it Chandra} data.
Our results are reported in this paper.
With {\it Chandra}'s fine angular resolution,
we confirm the location offset between the X-ray and radio nebulae of G16.73+0.08.
This morphology indicates that G16.73+0.08 is likely an evolved system interacting with the supernova-remnant
reverse shock \citep{blondin01,swaluw04,temim15,kolb17}.
We also 
identify
a point X-ray source near the peak of the X-ray nebula.
The spectrum of the X-ray nebula can be described by an absorbed power law, consistent with
earlier {\it XMM} observation.
We describe 
the {\it Chandra} data and image morphology in Section 2, spectral analysis in Section 3,
and discuss the location offset between the X-ray and radio nebulae of G16.73+0.08 and other
physical properties of the possible pulsar at the centre of the X-ray nebula in Section 4.

\section{Data and Image Morphology}

The {\it Chandra} X-ray Observatory \citep{weisskopf96} observed G16.7+0.1
on 2003 October 3 with ACIS-S without grating in a timed-exposure (TE) mode
(ObsId: 3845; PI: D. Helfand). The usable data volume, after GTI filtering, is about 25 ksec.
We processed this data using 
{\it Chandra} Interactive Analysis of Observations (CIAO)
software version 4.9
to produce level 2 data for further analysis.
We selected only those data  in the energy range from 1 keV to 8 keV to have better energy calibration information,
lower particle background and lower ACIS quantum efficiency contamination for this study.

\begin{figure}
\includegraphics[width=8.4cm]{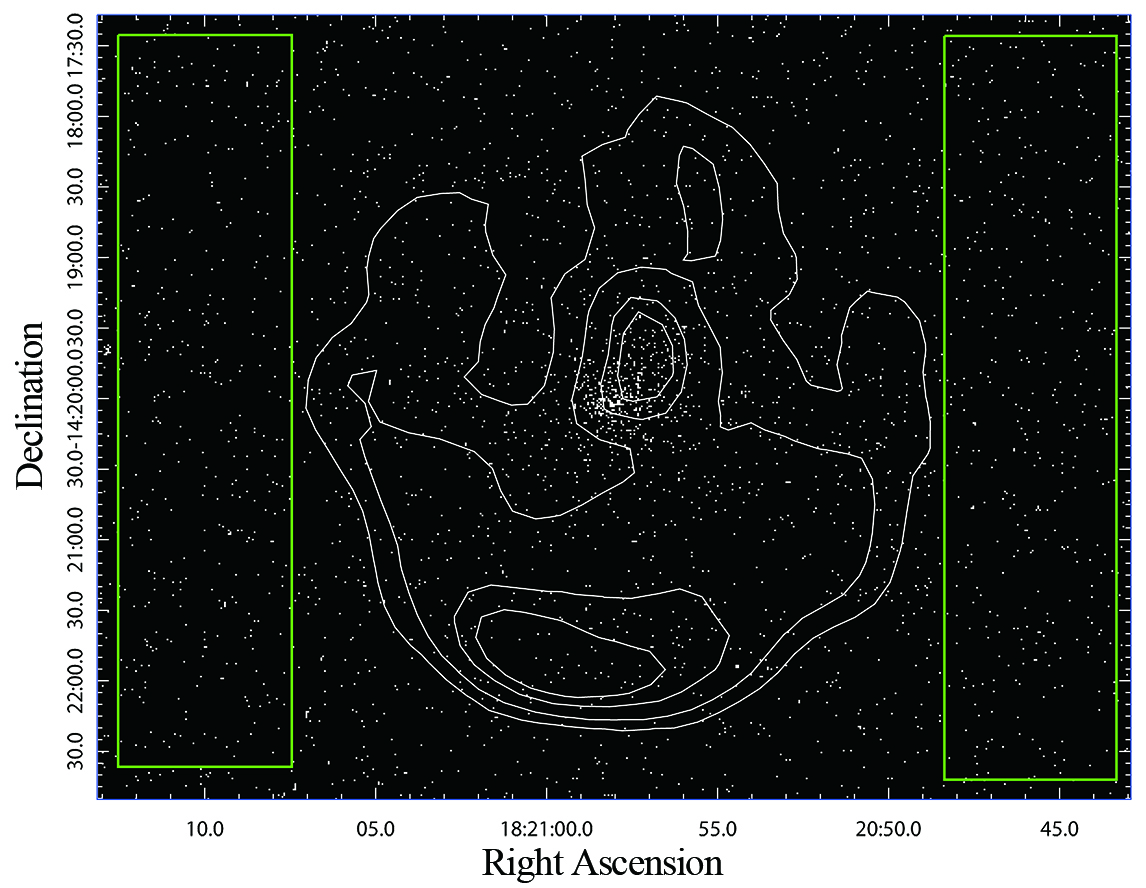}
\caption{{\it Chandra} image of G16.73+0.08 overlaid with a JVLA 1.4-GHz contour map of G16.7+0.1/G16.73+0.08.
The 1.4-GHz contours are at the levels of 10, 20, 35, and 45 mJy per beam, respectively.
The abscissa is the right ascension and the ordinate is the declination (Epoch 2000).
A zoom-in view of the central region is shown in Fig. \ref{zoomin}.
The two green rectangular boxes indicate the regions used for background estimate in our spectral analysis
described in Section 3.
}
\label{radioxray}
\end{figure}

\begin{figure}
\includegraphics[width=8.8cm]{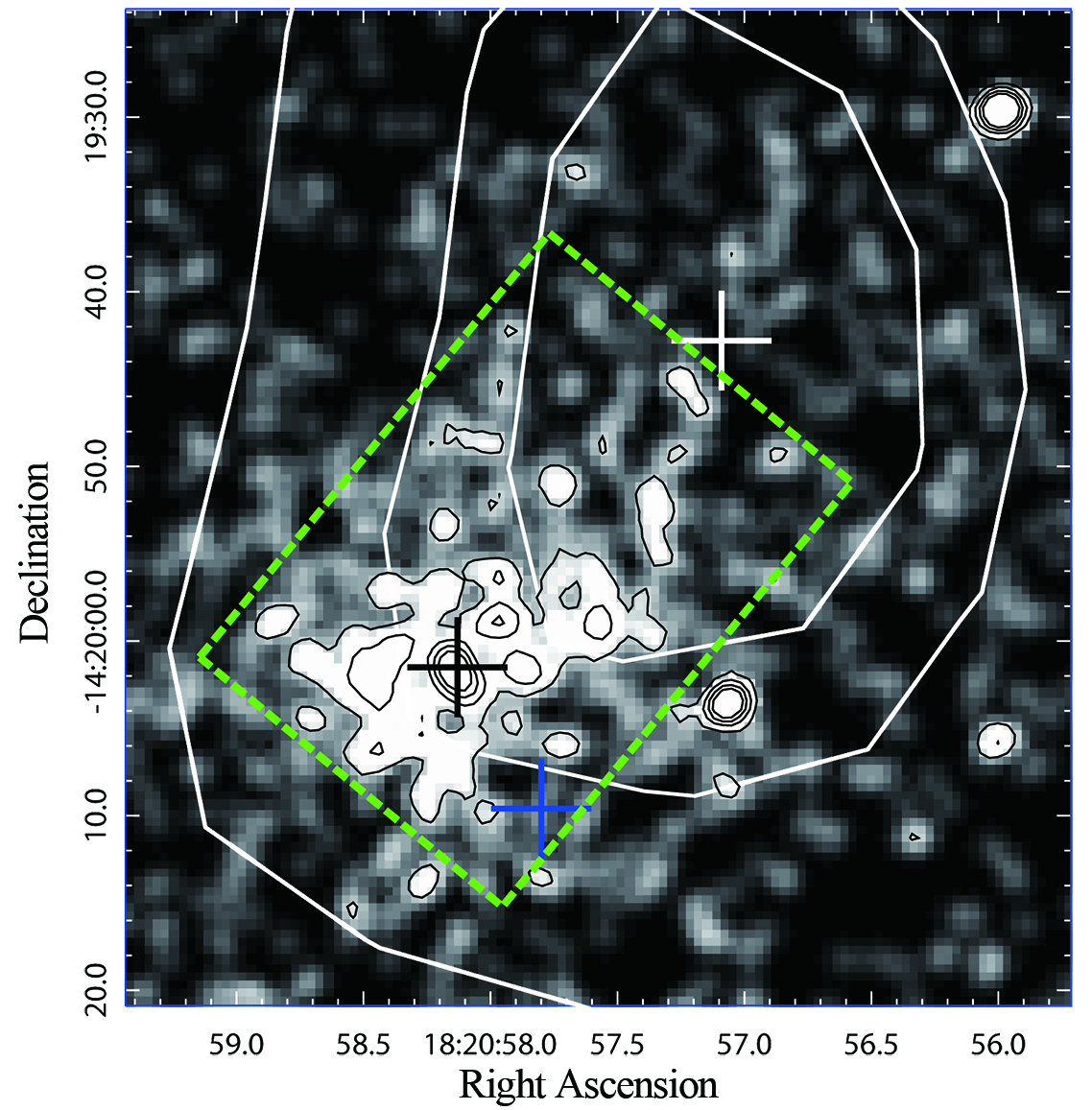}
\caption{A zoom-in view of the central region of Fig. \ref{radioxray}.
The {\it Chandra} image has been smoothed with a Gaussian filter of $\sigma=1.5$ pixels. 
The smoothed intensity 
is displayed with a logarithmic scale and is described quantitatively with black contours.
These X-ray contours are at the level of 0.25, 0.5, 0.75, and 1 count per pixel in the smoothed image.
The 1.4-GHz contours are the same as that in Fig. \ref{radioxray}.
The three crosses indicate the location of the X-ray emission peak (black), 
the centre of the 1.4-GHz core (white), and the geometric centre of G16.7+0.1 at 6 cm (blue).
The green rectangle encloses the target region for spectral analysis.
The two point sources, one at 16 arcsec to the west of black cross and the other
at the top right corner of this image, both have counterparts in the 2MASS point-source catalog.
}
\label{zoomin}
\end{figure}
Fig. \ref{radioxray} shows the {\it Chandra} image of G16.73+0.08 overlaid with JVLA 1.4-GHz continuum contours (see also\citet{zhang17}).
It is clear in this figure that the X-ray nebula does not coincident well with the 1.4-GHz nebula near the centre of
SNR G16.7+0.1.
Fig. \ref{zoomin} is a zoom-in view of this central region.
The three crosses indicate the positions of the X-ray emission peak 
(black, $\alpha=18^{\rm h}20^{\rm m}58^{\rm s}.13,\,\delta=-14^\circ 20' 01''.5$),
the centre of the 1.4-GHz nebula 
(white, $\alpha=18^{\rm h}20^{\rm m}57^{\rm s}.09,\,\delta=-14^\circ 19' 42''.8$), 
and the geometric centre of the radio SNR G16.7+0.1 
(blue, $\alpha=18^{\rm h}20^{\rm m}57^{\rm s}.8,\,\delta=-14^\circ 20' 09''.6$, as quoted in \citet{helfand03}).
The X-ray nebula is elongated in the direction from the X-ray peak to the 1.4-GHz nebula centre, with
a size of about 30 arcsec $\times$ 20 arcsec. 
Two point-like sources in Fig. \ref{zoomin}, located at 16 arcsec to the west of the X-ray peak and
at the top-right in the figure, are 
spatially coincident with two point sources in the 2MASS catalog,
2MASS 18205707-1420034 ($\alpha=18^{\rm h}20^{\rm m}57^{\rm s}.071,\,\delta=-14^\circ 20' 03''.49$)
and 2MASS 18205669-1419295 ($\alpha=18^{\rm h}20^{\rm m}56^{\rm s}.690,\,\delta=-14^\circ 19' 29''.58$), 
respectively. 

\begin{figure*}
\includegraphics[width=16cm]{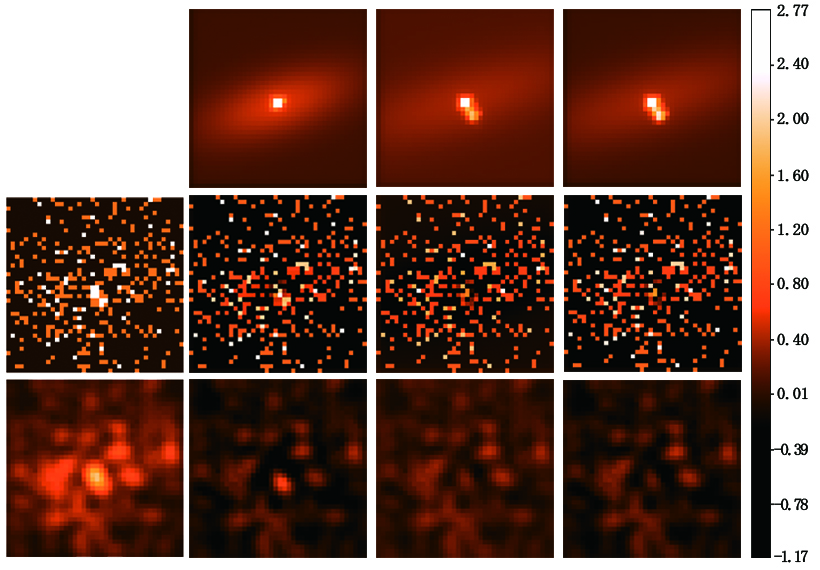}
\caption{Image morphology fitting in the 20-srcsec $\times$ 20-arcsec square area centred at the X-ray emission peak
location (black cross in Fig. \ref{zoomin}). 
The first column (counted from the left) shows the original {\it Chandra} image in that area.
The bottom panels in all columns are the corresponding middle-panel images smoothed with a Gaussian filter of $\sigma=1.5$ pixels. 
In the 2nd, 3rd and 4th columns, the top panels are images of different fitting models convolved with the {\it Chandra} point spread function, and the middle panels are residual images for that fitting model. See the main text for details of components employed in
different models. 
}
\label{imfit}
\end{figure*}

The X-ray emission peak, located at 24 arcsec away from the 1.4-GHz nebula centre to the southeast, is highly
concentrated.  
To study the morphology in more details, we used a model composed of a delta function, a two-dimensional Gaussian
and a constant 
to fit the image data in a 20-arcsec $\times$ 20-arcsec region centred at the X-ray emission peak location, the black cross
in Fig. \ref{zoomin}.
The model was convolved with a {\it Chandra} point spread function, which we obtained with the {\it Chandra} Ray Tracer (ChanRT)
simulator \citep{carter03} at photon energy 3.5 keV and with the MARX package version 5.3.2 \citep{davis12}. 
Results of this model fitting are shown in the 2nd column in Fig. \ref{imfit}. 
In the top panel is the convolved model.
The middle panel is the residual image and the bottom panel is the smoothed residual image.
Although this fitting is well acceptable with $\chi^2_\nu=0.19$, mainly due to the small number of photons, 397, in
this 40-pixel $\times$ 40-pixel region, a bright feature remains at the location of the X-ray emission peak.
We therefore added one more two-dimensional Gaussian to account for that feature.
Results of this fitting are  shown in the 3rd column in Fig. \ref{imfit}. 
This fitting is similarly acceptable, but that bright feature is gone. 
This second Gaussian component, whose FWHM is about 2.6 arcsec, might be manifestation of a possible torus around
the central neutron star represented here by the delta-function component.
Values of its fitting parameters are shown in Table \ref{imfitpa}.
We note, however, that a model without the delta-function component, that is, consisting of a constant and
two two-dimensional Gaussians, also gives an equally well fitting and there is no bright featute at the centre either.
Results of this fitting are  shown in the 4th column in Fig. \ref{imfit}. It casts some doubt on the existence
of a point source at the centre, as judged from the data. 
Nonetheless,  the model with a delta function and two Gaussians besides a constant, seems more physical, since it
presents a picture of a neutron star, a possible torus, and a pulsar wind nebula.
We 
identify
the central point source, represented here by the delta function component,
as CXO J182058.16-142001.5.

\begin{table}
\begin{center}
\begin{tabular}{ll}

\hline
\multicolumn{2}{l}{Model A (the 3rd column in Fig. \ref{imfit}): } \\
\hline
\multicolumn{2}{l}{\underline{The delta-function component}} \\
 location R.A.: & $18^{\rm h}20^{\rm m}58^{\rm s}.16$ \\
 location Dec.: & $-14^\circ 20' 1''.5$ \\
 photon count: & 12.32 \\
 & \\
\multicolumn{2}{l}{\underline{The 2-d Gaussian component I}} \\
 centre R.A.: & $18^{\rm h}20^{\rm m}58^{\rm s}.12$ \\
 centre Dec.: & $-14^\circ 20' 2''.4$ \\
 ellipticity: & 0.88 \\
 FWHM: & $2.56''$ \\
 major-axis direction: & $29.26^\circ$ from north to east \\
 photon count: & 13.96 \\
  & \\
\multicolumn{2}{l}{\underline{The 2-d Gaussian component II}} \\
 centre R.A.: & $18^{\rm h}20^{\rm m}58^{\rm s}.17$ \\
 centre Dec.: & $-14^\circ 20' 1''.2$ \\
 ellipticity: & 0.73 \\
 FWHM: & $48.61''$ \\
 major-axis direction: & $105.78^\circ$ from north to east \\
 photon count: & 123.48 \\
  & \\
\multicolumn{2}{l}{\underline{The constant component}} \\
 photon count: & 197.27 \\
\hline 

\hline
\multicolumn{2}{l}{Model B (the 4th column in Fig. \ref{imfit}): } \\
\hline
\multicolumn{2}{l}{\underline{The 2-d Gaussian component I}} \\
 centre R.A.: & $18^{\rm h}20^{\rm m}58^{\rm s}.13$ \\
 centre Dec.: & $-14^\circ 20' 2''.1$ \\
 ellipticity: & 0.88 \\
 FWHM: & $2.09''$ \\
 major-axis direction: & $32.75^\circ$ from north to east \\
 photon count: & 26.24 \\
  & \\
\multicolumn{2}{l}{\underline{The 2-d Gaussian component II}} \\
 centre R.A.: & $18^{\rm h}20^{\rm m}58^{\rm s}.17$ \\
 centre Dec.: & $-14^\circ 20' 1''.2$ \\
 ellipticity: & 0.72 \\
 FWHM: & $48.43''$ \\
 major-axis direction: & $105.79^\circ$ from north to east \\
 photon count: & 124.09 \\
  & \\
\multicolumn{2}{l}{\underline{The constant component}} \\
 photon count: & 194.84 \\
\hline
\end{tabular}
\end{center}
\caption{Best-fit parameter values of each component in the morphology model fitting.
Listed here are that of the two models shown in the 3rd and 4th column in Fig. \ref{imfit}.
The number of  photons in the fitting area (20 arcsec $\times$ 20 arcsec) is 397. 
The total model photon count is 347.03  for Model A and 345.17 for Model B.  
}
\label{imfitpa}
\end{table}
  
\section{Spectral Analysis}
To study the X-ray spectrum of this elongated nebula, we defined the target region for spectral analysis as the rectangular region shown in Fig. \ref{zoomin}. 
The background region was chosen to be the two rectangular regions outside the SNR, as shown in
Fig. \ref{radioxray}.
The spectrum of this target region, in $1 - 8$ keV, can be well fitted ($\chi^2_\nu=0.78$ for 22 d.o.f)
by 
an absorbed
power law of a photon index 
$\Gamma=0.99^{+0.68}_{-0.62}$ ($N_E\propto E^{-\Gamma}$) and a neutral hydrogen column density
$N_{\rm H}=5.16^{+2.28}_{-1.92}\times 10^{22}$ cm$^{-2}$.
The interstellar absorption model employed in our analysis is an photoelectric  absorption model (`phabs' in XSPEC)
with the assumed abundance taken from \citet{anders89}.
These results are consistent with the earlier XMM observation \citep{helfand03}.
On the other hand, we actually expect this region contains not only the pulsar wind nebula, but also some contribituion from
the SNR.  
Adding one more thermal spectral componet (`bremss' in XSPEC) to the model, 
however, yields a best fit with a larger $\chi^2$, 0.85. 
Although it is still an acceptable fit, the thermal component becomes the dominating one in such a fit,
and the uncertainty ranges of its fitting parameters are so large that XSPEC (version 12.0.9n)
cannot give a valid error estimate.
We do not think this is an adequate fit.
We conclude that the contribution from the SNR is negligible based on the current data.

\begin{figure}
\includegraphics[width=8.4cm]{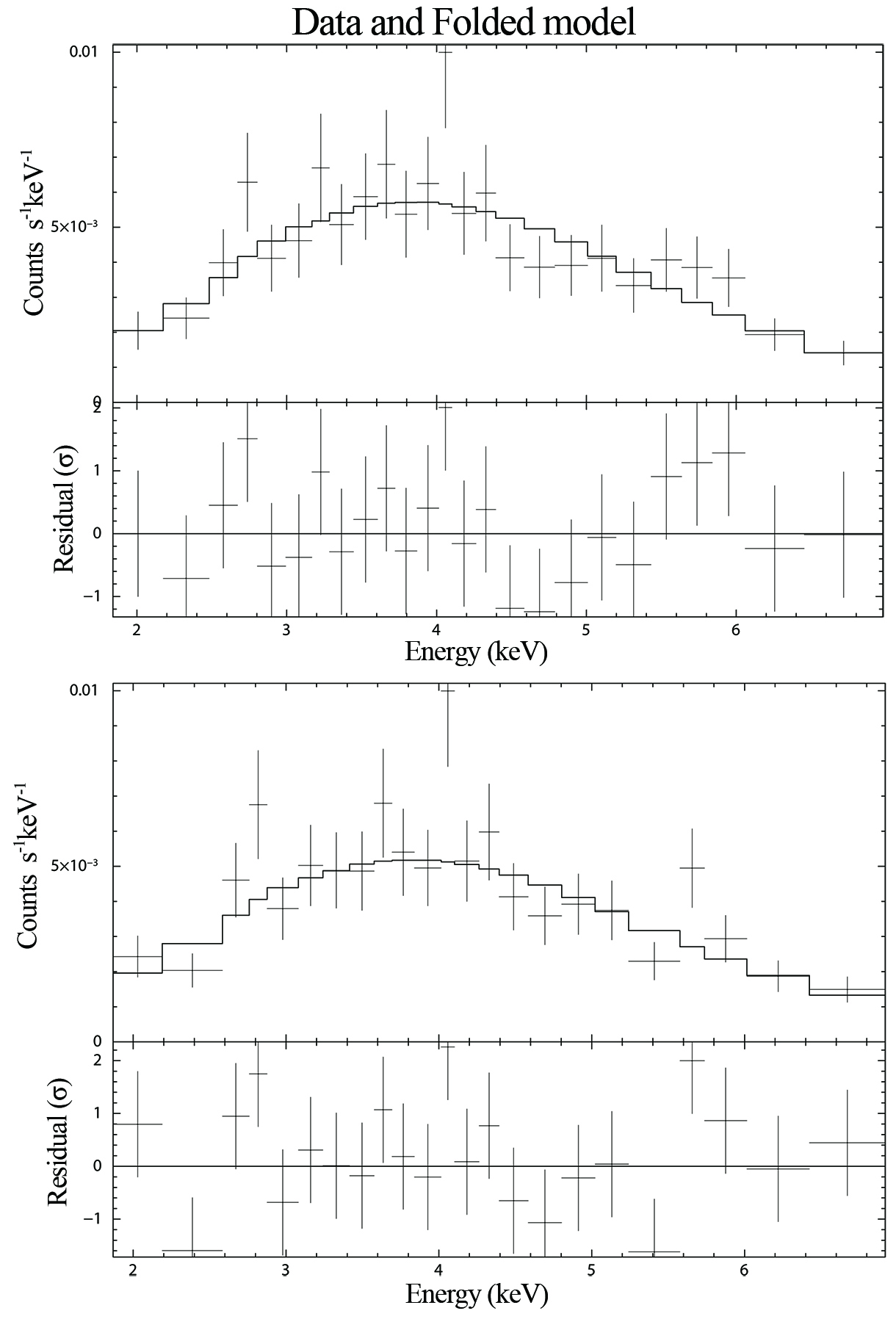}
\caption{Spectral fitting with {\it Chandra} ACIS-S data of the G16.73+0.08 X-ray nebula. 
The upper panel shows the spectrum of the whole target region, while the lower panel
shows that with the central circular region removed.
The model used is an absorbed power law. The best fit parameters are shown in Table \ref{spfitpa}.
}
\label{spfit}
\end{figure}

Although the above power-law fitting describes the spectrum very well, 
the central neutron star, together with the tiny extended source as revealed in our imaging analysis,
can have a  spectral behavior very different from the PWN. 
We therefore also performed another spectral analysis with the central region, defined as a circular region of radius 2 arcsec centred at
the X-ray peak location, removed from the PWN.
\begin{table}
\begin{center}
\begin{tabular}{ll}
\hline
\hline
 \multicolumn{2}{l}{\underline{The whole target region}} \\
Column density, $N_{\rm H}$ &  $5.16^{+2.28}_{-1.92}\times 10^{22}$ cm$^{-2}$ \\
Photon index, $\Gamma$ & $0.99^{+0.68}_{-0.62}$ \\
$\chi^2_\nu$ & 0.78 \\
Degrees of freedom & 22 \\
Unabsorbed flux (0.5 - 10 keV) & $1.3 \times 10^{-12}$ erg cm$^{-2}$ s$^{-1}$\\
 & \\
 \multicolumn{2}{l}{\underline{Target region without the central circle}} \\
Column density, $N_{\rm H}$ &  $4.99^{+2.75}_{-2.28}\times 10^{22}$ cm$^{-2}$ \\
Photon index, $\Gamma$ & $0.98^{+0.79}_{-0.71}$ \\
$\chi^2_\nu$ & 1.19 \\
Degrees of freedom & 20 \\
Unabsorbed flux (0.5 - 10 keV) & $1.2 \times 10^{-12}$ erg cm$^{-2}$ s$^{-1}$\\
\hline
\hline
\end{tabular}
\end{center}
\caption{Spectral fitting results. 
}
\label{spfitpa}
\end{table}
We fit the spectrum of the PWN target region without the central circle with a power-law model.
Fitting results are shown in Table \ref{spfitpa}. 
The neutral hydrogen column density, $4.99^{+2.75}_{-2.28}\times 10^{22}$ cm$^{-2}$
and the photon index, $\Gamma=0.98^{+0.79}_{-0.71}$, do not change much.
Similar to the spectrum of the whole target region with the central circle included, adding a thermal component
to the spectral model does not improve the fitting.
The fitting and residuals are shown in Fig. \ref{spfit}. 

The central circle contains only 39 photons in $1 - 8$ keV, out of the 512 photons in the whole target region. 
It is difficult to have a meaningful spectral fitting.
Its photon count is 11, 11, and 17 in the energy bands 1 -- 3.4, 3.4 -- 4.7, and 4.7 -- 8.0 keV, respectively.
The corresponding number for the whole target region is 164, 180, and 168.

In the {\it Chandra} X-ray image (Fig. \ref{radioxray}), the X-ray emission in the SNR shell region
is not obvious. Taking an annular region centred at the geometric centre of the radio SNR (the blue cross
in Fig. \ref{zoomin}) with a 45-arcsec inner radius and a 135-arcsec outer radius as the SNR shell region, the same as that defined
in \citet{helfand03}, and another annular region from 135 arcsec to 175 arcsec as
the background region, we found that there are 13014 photons in the SNR shell region and 
9391 counts in the background region.
The number of net counts in the shell region is 748.6, after subtracting estimated background counts.
This number is 6.5 times the square root of 13014.
We therefore conclude that the X-ray emission in the SNR shell region
is detected at a level of about $6.5\sigma$. This region is so background-dominated that a meaningful
spectral fitting cannot be obtained.

\section{Discussion}
 
The offset between the 1.4-GHz and X-ray nebulae in G16.73+0.08 
was revealed to some extent in the {\it XMM} data \citep{helfand03}, in which,
with a coarser spatial resolution, the X-ray nebula was found to be close to the
geometric centre of the radio SNR, while its radio nebula was known to be off centre \citep{helfand89}.
With {\it Chandra}'s better resolution, we report the offset of 24 arcsec between the 1.4-GHz and X-ray nebulae in G16.73+0.08
in Fig. \ref{zoomin}.
We note that the nebula observed at 6-cm and 20-cm wavelengths has a weak extension to the location of the X-ray nebula and that the 2-cm emission in fact has a high intensity region surrounding the location of CXO J182058.16-142001.5, in addition to the 2-cm nebula coincident with that at 6-cm, 20-cm, and 1.4-GHz (Figure 2 in \citet{helfand89}).
Such an offset between radio and X-ray nebulae has been observed in several pulsar wind nebulae, for example,
G327.1-1.1 \citep{temim15,temim09},
G292.0+1.8 \citep{bhalerao15}, 
MSH 15-56 \citep{temim13}, 
MSH 11-62 \citep{slane12}, 
and Vela X \citep{lamassa08}.
The morphology of these PWNe in X-ray and radio bands can be 
interpreted as PWNe being interacting with (or, crushed by) the supernova remnant reverse shock
\citep{blondin01,swaluw04,temim15,kolb17}.
The asymmetry of the SNR reverse-shock sweeping direction may come from inhomogeneity of ISM,
anisotropic SN explosion, and/or the motion of pulsars.
Our study indicates that G16.73+0.08 is a member of this reverse-shock-crushed PWN family. 
The age of SNR G16.7+0.1 
could therefore be larger than previuosly thought, 
because in such a case the SNR has passed the free expansion phase and is likely dynamically evolved.
Though, its absolute age could still be on the order of a few thousand years if the ambient density is high.
It depends strongly on the environment property; see, e.g., \citet{temim15} and \citet{kolb17}.

From the spectral analysis of the X-ray nebula, we obtained a photon index of $0.99^{+0.68}_{-0.62}$ 
and column density $N_{\rm H}=5.16^{+2.28}_{-1.92}\times 10^{22}$ cm$^{-2}$,
which are both in agreement with the photon index $1.17\pm0.29$ and 
$N_{\rm H}=4.74\pm 0.98\times 10^{22}$ cm$^{-2}$ obtained from {\it XMM} observation \citep{helfand03}.
The unabsorbed flux in 0.5 -- 10 keV from our result is
$1.3\times 10^{-12}$ erg s$^{-1}$ cm$^{-2}$, comparable to the flux $1.9\times 10^{-12}$ erg s$^{-1}$ cm$^{-2}$
obtained in \citet{helfand03}.
Removing the contribution from the central source in a 2-arcsec-radius circular region does not change
the fitting results much; see Table \ref{spfitpa}.

Based on an empirical relationship
between
the photon index of the X-ray nebula and the central pulsar's
spin-down power \citep{gotthelf03}, the photon index 0.99 gives a
spin-down power of $\dot{E}=2.2\times 10 ^{36}$ erg s$^{-1}$ for
the possible pulsar in G16.73+0.08.
Using another empirical relationship
between the X-ray luminosity $L_X$ (0.2 –- 4 keV) of the pulsar-powered nenula (including the
pulsar) and the pulsar spin-down power $\dot{E}$, 
$\log L_X = 1.39 \log\dot{E}-16.6$ \citep{seward88},
we may also estimate the spin-down power.
The derived unabsorbed flux from our fitting model in 0.2 -- 4 keV is
$5.1\times 10^{-13}$ erg s$^{-1}$ cm$^{-2}$.
Adopting a distance of 14 kpc \citep{zhang17}, the corresponding luminosity is
$1.2\times 10^{34}$ erg s$^{-1}$, which yields a spin-down power of
$\dot{E}=2.9\times 10 ^{36}$ erg s$^{-1}$. 
We will use the average of the above two estimates, that is, 
$\dot{E}=2.6\times 10 ^{36}$ erg s$^{-1}$, for the following discussion.

In \citet{helfand03}, the age of the possible pulsar is estimated to be about 2100 years,
based on the assumption that the SNR is in the free-expansion phase with an expansion
speed of 3000 km/s and the SNR is at a distance of 10 kpc. 
Using a new distance measurement based on HI absorption and OH maser observations \citep{zhang17}, 
the distance to G16.7+0.1 is inferred to be 14 kpc.
The age of the pulsar is then about 3000 years.
However, considering the scenario that SNR reverse shock has already crushed the PWN, 
its age 
could be 
larger.
Given the spin-down power $\dot{E}$ and characteristic age $\tau$, 
one may infer the pulsar's period $P$ and its time derivative $\dot{P}$.
For $\tau=3000$ years, we have $P=0.29$ s and $\dot{P}=1.5\times 10^{-12}$ s s$^{-1}$.
For $\tau=10000$ years, they are $P=0.16$ s and $\dot{P}=2.5\times 10^{-13}$ s s$^{-1}$.
The estimated dipole magnetic field strength at the polar surface is 
$4.2\times 10^{13}$ G and $1.3\times 10^{13}$ G for the above two cases, respectively.

In summary, the PWN G16.73+0.08 has its X-ray emission region separated from the radio one.
It is likely that this PWN is in the stage of interacting with the SNR reverse shock.
A point-like X-ray source, which we 
designate
as CXO J182058.16-142001.5, 
is found near the brightness centre of the X-ray nebula, together with a small extended source,
which may be a torus surrounding the neutron star. 
Future longer observations with adequate time resolution is very much desired to explore the
properties of this possible pulsar. Observations with adequate spatial resolution 
is also needed to study spatially resolved spectra of the nebula. 
These are helpful to our understanding of how pulsars emit electron-positron pairs
and how these pair plasmas, as the pulsar winds, propagate and evolve. 

\section*{Acknowledgments}
We are very much appreciative of valuable suggestions from the anonymous referee to
improve this paper significantly. This research has made use of data obtained through 
the High Energy Astrophysics Science Archive Research centre Online Service, 
provided by the NASA/Goddard Space Flight Center. 
This work was supported by the Ministry of Science and Technology of 
the Republic of China under grants MOST 105-2112-M-007-002 and MOST 106-2811-M-007-008.

\label{lastpage}
\end{document}